\documentclass[aps,twocolumn,prl,tightenlines,floatfix,showpacs]{revtex4}

\usepackage[dvips]{graphicx}
\usepackage[english]{babel}
\usepackage{amsmath}
\usepackage{amssymb}
\usepackage{times}

\begin{document}

\newcommand{\D}{\mathrm{D}}
\newcommand{\p}{\partial}
\newcommand{\Tr}{\mathrm{Tr}}

\title{Finite-Temperature Behavior of an Inter-species Fermionic Superfluid
with Population Imbalance}

\author{Hao Guo$^{1}$, Chih-Chun Chien$^{1}$, Qijin Chen$^{2,1}$, 
Yan He$^{1}$, and K. Levin$^{1}$}

\affiliation{$^1$James Franck Institute and Department of Physics,
University of Chicago, Chicago, Illinois 60637, USA}
\affiliation{$^2$Zhejiang Institute of Modern Physics and Department of
Physics, Zhejiang University, Hangzhou, Zhejiang 310027, China}

\date{\today}

\begin{abstract}
  We determine the superfluid transition temperature $T_c$ and related
  finite temperature phase diagrams for the entire BCS-Bose Einstein
  condensation crossover in a homogeneous mixture of $^{6}$Li and
  $^{40}$K atoms with population imbalance.  Our work is motivated by
  the recent observation of an inter-species Feshbach resonance.
  Pairing fluctuation effects, which significantly reduce $T_c$ from the
  onset temperature for pairing ($T^*$), provide reasonable estimates of
  $T_c$ and indicate that the inter-species superfluid phase should be
  accessible in future experiments. Although a generalized-Sarma
  phase is not stable in the ground state near unitarity, our phase
  diagrams show that it appears as an intermediate-temperature
  superfluid.  
\end{abstract}

\pacs{03.75.Ss,03.75.Hh,05.30.Fk}

\maketitle

Ultracold Fermi gases with tunable attractive interactions provide an
exciting opportunity to study superfluidity in the very general context
of a crossover from BCS theory to Bose-Einstein condensation
(BEC). While initial experiments addressed the more conventional
situation of intra-species pairing with equal populations of the two
effective spin states, more recently there has been an emphasis on
population imbalanced gases
\cite{ZSSK06,MITPRL06,Rice1,Rice2-old,KetterleRF}. Many interesting
phases have been contemplated including two which appear to have been
observed in the laboratory: the so-called Sarma phase
\cite{Sarma63,our06} which represents a homogeneously polarized state,
and a phase separated (or inhomogeneously polarized) state
\cite{Caldas03_04}.
%, as well as the yet-to-be-discovered
%Larkin-Ovchinnikov-Fulde-Ferrell (LOFF) phase 
%\cite{FFLO,LOFF1}.

Adding to the excitement is the possibility of discovering a new form of
superfluid involving inter-species pairing.  A first step en route to
the discovery is the recent observation of Feshbach resonances between
$^{6}$Li and $^{40}$K atoms \cite{Wille08}. If the transition
temperatures are accessible, this tunable attractive interaction should
enable BCS-BEC crossover in superfluid phases associated with unequal
mass pairing.
Moreover, it is of particular interest for its relevance to color
superconductivity in quark matter \cite{Nardulli}.

In this paper we determine the transition temperatures for the entire
BCS-BEC crossover in a homogeneous mixture of $^{6}$Li and $^{40}$K
atoms with population imbalance.  In addition we address the
temperature-polarization phase diagrams associated with inter-species
superfluid phases at and around unitarity.  We consider only a
homogeneously polarized superfluid and exclude from consideration the
phase separated state principally because it is now clear
\cite{KetterleRF} that the normal regions in this heterogeneous phase
correspond to a complicated correlated normal state, which is currently
difficult to characterize at an analytical level,
except in the limit of extreme population imbalance.  
% can do it at p=0.44 with QMC.
This normal state, which is distinct from a free Fermi gas, has been addressed
numerically using
quantum Monte Carlo
simulations \cite{GiorginiPS}.

Our $T \neq 0$ calculations are performed in a fashion consistent with a
generalized BCS-Leggett ground state which has been
studied previously \cite{IsMe,ParishPRL07} in the strict $T=0$ limit.
These calculations show the importance of including pairing fluctuations
which greatly suppress the transition temperature $T_c$ from the pair
formation temperature $T^*$.  An understanding of finite temperature
effects positions us to address actual experiments (which are never in
the ground state).  Moreover, new superfluid phases appear which are
constrained to an intermediate regime of non-zero temperatures. That is,
they are associated with a lower as well as upper critical temperature.
We show that, in the absence of a trap,
this intermediate temperature superfluid will be extremely  
difficult to observe when the heavy species is the majority, but it
should be more accessible for the case where the heavy species is the minority.
Finally, we study how the phase diagram evolves as one crosses from BCS
to BEC.  In contrast to polarized Fermi gases with equal mass, close to,
but on the BEC side of resonance, the intermediate temperature
superfluid disappears when the lighter species is the majority,
%for the appropriate polarization sign,
giving way to a conventional polarized superfluid with only one
transition temperature.

There have been extensive studies on zero temperature properties of
homogeneous \cite{IsMe} as well as trapped inter-species Fermi gases 
\cite{IsMe,Yip2mass,YD05}.
This body of work (like that in the present paper) is based on a natural
generalization of the BCS-Leggett ground state to
accommodate unequal populations.  Similarly,
while a two channel model formalism may be more relevant to the narrow
resonances seen in Ref. \cite{Wille08}, all work to date (including our
own) deals with the simpler one channel model.  Theoretical studies at
finite temperatures that are
%, designed to be 
consistent with these $T=0$
calculations has been limited to a strict mean field approach
\cite{ParishPRL07} which ignores the important effects of pair
fluctuations, or non-condensed pairs.

We begin with an outline of the central equations associated with a
generalized ``Sarma'' state appropriate to the case of unequal masses
and finite polarization. This is followed by a short microscopic
derivation.  We choose the convention $m_{\downarrow}>m_{\uparrow}$,
so that the mass of spin-down fermions is the larger. The mass ratio is
$m_{\downarrow}/m_{\uparrow}=6.7$.  We define
$E_{k\uparrow,\downarrow}=E_{k}\pm\xi^{-}_{k}$,
$E_{k}=\sqrt{\xi_{k}^{+2}+\Delta^{2}}$, where
$\xi^{\pm}_{k}=(\xi_{k\uparrow}\pm\xi_{k\downarrow})/2$.
Here 
$\xi_{k\sigma}=\epsilon_{k\sigma}-\mu_{\sigma}$, and
$\epsilon_{k\sigma}=k^{2}/2m_{\sigma}$, where $\sigma=\uparrow,
\downarrow$.  The four unknowns which must be determined at general
temperature $T$ involve the two fermionic chemical potentials
$\mu_{\uparrow}$, $\mu_{\downarrow}$, and the excitation gap $\Delta$,
whose square appears in $E_{k}$. The quantity $\Delta^2$ contains a
contribution from condensed (sc) and non-condensed (pg) pairs:
\begin{equation}
\Delta^{2}=\Delta^{2}_{sc}+\Delta^{2}_{pg}
\label{eq:1z}
\end{equation}
so that one of the gap components ($\Delta_{pg}$) must be separately
determined in order to establish the transition temperature $T_c$. This
is the lowest temperature at which $\Delta_{sc}$ vanishes.

There are then four equations, three of which can be associated with
strict mean field theory, and all of which are derived microscopically
below. The equations for the total number
$n=n_{\uparrow}+n_{\downarrow}$ and number difference $\delta
n=n_{\downarrow}-n_{\uparrow}$ of fermions are
\begin{eqnarray}\label{eq:ntot} 
n
%&=&n_{\uparrow}+n_{\downarrow}
%\nonumber \\
&=&\sum_{\mathbf{k}}\Big\{\Big(1-\frac{\xi_{k}^{+}}{E_{k}}\Big)+[f(E_{k\uparrow})+f(E_{k\downarrow})]\frac{\xi_{k}^{+}}{E_{k}}\Big\}
\end{eqnarray}
and
\begin{equation}\label{eq:dn}
\delta
n=\sum_{\mathbf{k}}[f(E_{k\downarrow})-f(E_{k\uparrow})].
\end{equation} 
Here $f(x)=(e^{x/T}+1)^{-1}$ is the Fermi distribution function.  The
gap parameter $\Delta$ is obtained from
\begin{equation}\label{eq:geq} -\frac{M}{2\pi
    a}=\sum_{\mathbf{k}}\Big[\frac{1-f(E_{k\uparrow})-f(E_{k\downarrow})}{2E_{k}}-\frac{1}{\epsilon_{k}}
  \Big]. \end{equation} Here the coupling constant is regularized by
$g^{-1}=M/(2\pi a)-\sum_{\mathbf{k}}(2\epsilon_{k})^{-1}$, where $a$ is
the s-wave scattering length,
$M=m_{\uparrow}m_{\downarrow}/(m_{\uparrow}+m_{\downarrow})$ is the
reduced mass and $\epsilon_{k}=k^{2}/2M$.

Finally, we need an equation for $\Delta_{pg}^2$ which requires that we
establish the dispersion of the non-condensed pairs.  These
non-condensed pairs, or pseudogap effects, appear at $T \neq 0$ and are
included via a $T$-matrix contribution to the fermion
self-energy. Following Refs.~\cite{ourreview,our06}, the fermionic self
energy $\Sigma_{\sigma}(K)=\sum_{Q}t(Q)G_{\bar{\sigma}}(Q-K)$, where the
four-momentum $Q=(i\Omega_{l},\mathbf{q})$, $K=(i\omega_n, \mathbf{k})$,
and $\Omega_{l}$ ( $\omega_n$ ) is the bosonic (fermionic) Matsubara
frequency, with $\sum_{Q}=T\sum_{l}\sum_{\mathbf{q}}$,
$\sum_{K}=T\sum_{n}\sum_{\mathbf{k}}$, and $\bar{\sigma}=-\sigma$. The
$T$-matrix is presumed to have the structure
$t(Q)=t_{sc}(Q)+t_{pg}(Q)$. The condensate contribution satisfies
$t_{sc}(Q)=-(\Delta_{sc}^{2}/T)\delta(Q)$.  Here the fermionic Green's
function is
$G_{\sigma}(K)=[G^{-1}_{0\sigma}(K)-\Sigma_{\sigma}(K)]^{-1}$, with
$G^{-1}_{0\sigma}(K)=(i\omega_n -\xi_{k\sigma})$.  We
%define $K=(i\omega_n, \mathbf{k})$, and
set $\hbar\equiv 1$ and $k_{B}\equiv 1$.
%$\omega_n$ is the fermion Matsubara frequency.

The excited pair propagator (which is what we need to arrive at our
fourth equation) is given by $t_{pg}(Q)=[g^{-1}+\chi(Q)]^{-1}$, where
the symmetrized pair susceptibility, $\chi(Q) =
\sum_{K,\sigma}G_{0\sigma}(Q-K)G_{\bar{\sigma}}(K)/2$, is used.  
% Here $\sum_{K}=T\sum_{n}\sum_{\mathbf{k}}$.
A central assumption is the usual BEC condition that the pair chemical
potential vanishes below $T_c$, which will, in turn, lead to
Eq.~(\ref{eq:geq}).  This BEC condition implies that $t_{pg}$ is
dominated by terms with $Q\approx 0$. Importantly, the pseudogap is determined by
%component to the gap
$\Delta^{2}_{pg}\equiv -\sum_{Q}t_{pg}(Q)$. It follows that $\Sigma_{\sigma}(K) =
-\Delta^{2}G_{0\bar{\sigma}}(-K)$, and in that way we have derived
Eq.~(\ref{eq:1z}). One arrives at the two
equations for the number densities via
$n_{\sigma}=\sum_{K}G_{\sigma}(K)$ and in this way derives
Eqs. ~(\ref{eq:ntot}) and ~(\ref{eq:dn}).

The $T$-matrix may be expanded \cite{expan} as
$t_{pg}^{-1}(Q)=a_{0}\Omega+a_{1}\Omega^{2}-\xi^{2}q^{2}$, after
analytic continuation ($i\Omega_{l}\rightarrow\Omega+i0^+$), where we
have neglected the small imaginary part $\Gamma_Q$. The pseudogap
contribution can be written as
\begin{equation}
\Delta_{pg}^{2}=\sum_{\mathbf{q}}\frac{b(\tilde{\Omega}_{\mathbf{q}})}{\sqrt{a_{0}^{2}+4a_{1}\xi^{2}q^{2}}}\,.
\end{equation} 
Here $b(x)$ is the Bose distribution function and
$\tilde{\Omega}_{\mathbf{q}} =
(\sqrt{a_{0}^{2}+4a_{1}\xi^{2}q^{2}}-a_{0})/2a_{1}$.  In the BEC limit,
it can be shown that $a_{1}/a_{0}\rightarrow 0$ and
$\Omega_{\mathbf{q}}\rightarrow q^{2}/2M^{*}$, where
$M^{*}=a_{0}/2\xi^{2}$ is the effective pair mass. Importantly, in this
limit $M^{*}$ approaches the total mass of the two constituent fermions,
and $T_c$ approaches the BEC temperature of ideal bosons of
density $ \min(n_\uparrow, n_\downarrow)/2$ and mass $ M^*$.

The gap equation, Eq.~(\ref{eq:geq}), is equivalent to an extremal
condition on the thermodynamic potential
$\partial\Omega_{MF}/\partial\Delta=0$, where
\begin{equation} 
  \Omega_{MF}=-\frac{\Delta^{2}}{g}+ 
  \sum_{\mathbf{k}}(\xi^{+}_{k}-E_{k})- T\sum_{\mathbf{k},\sigma}\Big[\ln
  \left(1+e^{-E_{k\sigma}/T}\right)\Big]. 
\end{equation}
Superfluid stability requires that the number susceptibility matrix
$\partial n_{\sigma}/\partial\mu_{\sigma^{\prime}}$ should be
positive-definite \cite{our06}. This can be shown to coincide
% Because every element of the matrix has a common factor
% $\partial^{2}\Omega_{MF}/\partial\Delta^{2}$ in the denominator, the
% eigenvalues of the number susceptibility matrix change sign when
% $\partial^{2}\Omega_{MF}/\partial\Delta^{2}$ does.
with the condition that $\partial^{2}\Omega_{MF}/\partial\Delta^{2}>0$. 
When this condition is violated, phase separation may occur.  These
alternative phases have been widely studied \cite{SR06} at $T=0$, as
well as at non-zero temperature \cite{ChienPRL}.

\begin{figure} 
\includegraphics[width=3.4in,clip] {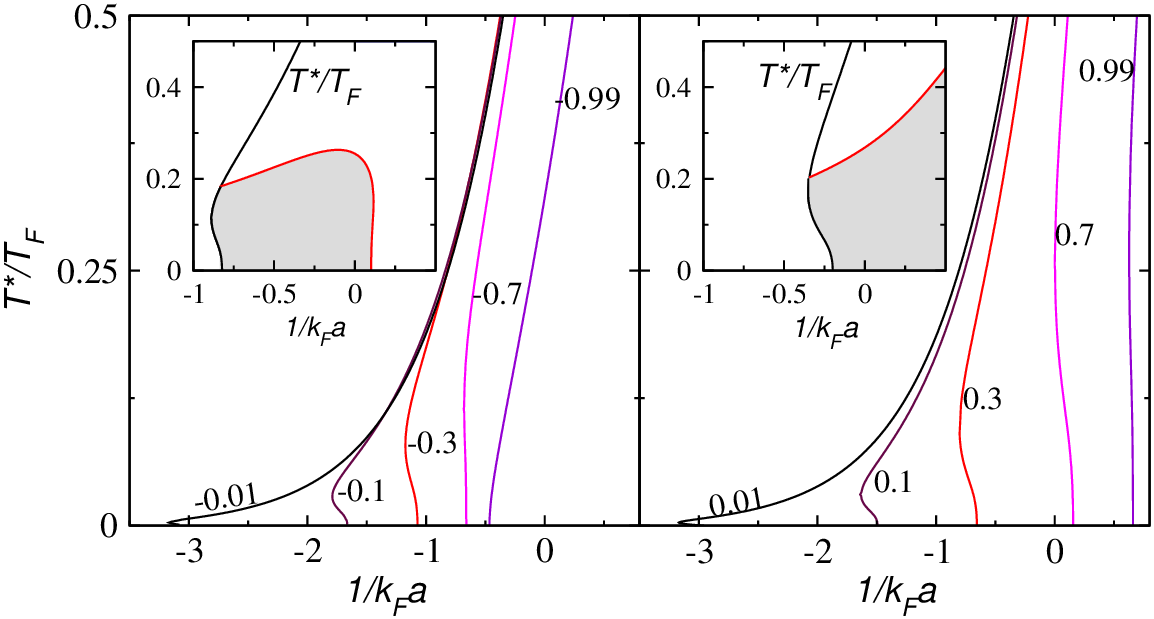} 
\caption{(Color online) $T^{*}$ as a function of $1/(k_{F}a)$ for
  several values of $p$ (as labelled).  Here $p<0$ when the lighter
  (spin up) species is the majority (left panel) and $p>0$ otherwise.
  Insets: $T^*$ (Black solid line) and unstable regime (shaded regime)
  for $p=-0.5$ and $p=0.5$, respectively.}
\label{fig:TS} 
\end{figure}

The distinction between $\Delta$ and the order parameter $\Delta_{sc}$
is an important component of the present theory. The former is
associated with an onset temperature $T^*$ and the latter with $T_c$.
We arrive at a reasonable estimate of $T^*$ from
Eqs.~(\ref{eq:ntot})-(\ref{eq:geq}) which is plotted in
Fig.~\ref{fig:TS} as a function of $1/(k_{F}a)$.  Two different signs of
the polarization $p = \delta n/n$ are indicated in the right and left
panels.  Here $k_{F}$ and $T_{F}$ are the Fermi momentum and the Fermi
temperature of an unpolarized non-interacting Fermi gas with the same
total particle density presuming a mass equal to the average mass of $^{6}$Li
and $^{40}$K. We will see that throughout the paper
% It can be shown that at $T=0$ when polarization is small in the BEC
% regime,
a superfluid phase with $p<0$ (the lighter species is
the majority) appears to be more readily obtainable than one with
$p>0$. The arguments behind this asymmetry in $p$ are subtle and involve
both energetic comparisons as well as mechanical stability (in the sense
of $\partial^{2}\Omega_{MF}/\partial\Delta^{2}>0$).  It can be
analytically shown that at $T=0$ when polarization is small in the BEC
regime, a superfluid phase with $p<0$
% (corresponding to the lighter species as the majority)
is more energetically favorable than one with $p>0$.

% The behavior of $T^*$ is similar to that shown in Fig.1 of
% Ref.~\cite{our06} for polarized Fermi gases with equal mass.
The figure shows that $T^*$ vanishes when the attraction is sufficiently
weak and near its vanishing point it displays non-monotonic
behavior. Similar behavior has been observed previously for population
imbalanced Fermi gases of equal masses \cite{our06}. The insets of
Fig.~\ref{fig:TS} indicate the unstable regimes for $p=\pm 0.5$. The
unstable regimes are asymmetric in the sign of $p$ and at $T=0$ the
behavior is
% because the unstable regime extends to higher temperature and into
% deeper BEC regime when the heavier fermions are the majority. In Fermi
% gases with equal mass no such asymmetry was observed. This observation
% is
consistent with phase diagrams obtained earlier \cite{IsMe,ParishPRL07}.

\begin{figure} 
\includegraphics[width=3.4in,clip] {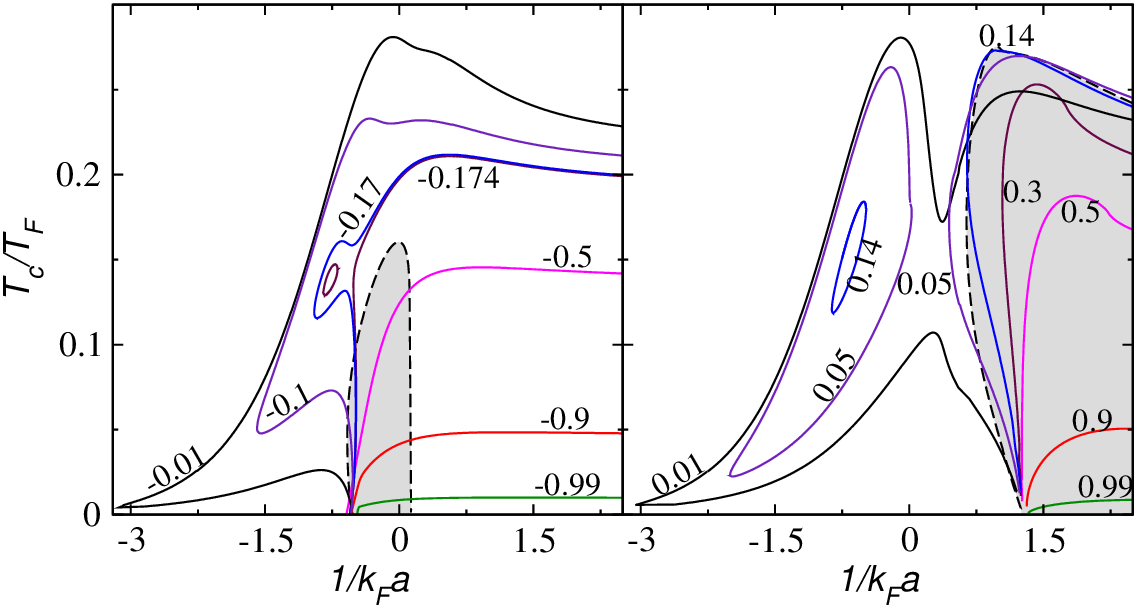} 
\caption{(Color online) $T_{c}$ as a function of $1/(k_{F}a)$ for
  selected values of $p$. Here $p<0$ when the lighter species is the
  majority (left panel) and $p>0$ otherwise. The polarized superfluid
  solution is unstable in the shaded regions.}
\label{fig:Tc} 
\end{figure} 

We turn next to the superfluid transition temperature, $T_c$, which is
plotted in Fig.~\ref{fig:Tc} as a function of $1/(k_{F}a)$, for both
$p<0$ (left panel) and $p>0$ (right panel).
% Here the left and right panels correspond to negative and positive
% values of $p$, respectively.  This is to be compared with the
% equal-mass case shown in Fig.2 of Ref.~\cite{our06}.
It is clear that these homogeneous superfluid phases are more likely to
be observed when polarization is low. The shaded regions
% in the figure
indicate where this form of superfluidity is unstable. Note that there
is a rather pronounced asymmetry between the $p < 0$ and $ p >0$ cases.
Indeed, when $p>0$, a stable superfluid cannot be found near $1/k_F
a=1.5$ although it will emerge again deep in BEC.
An ``intermediate temperature superfluid'' phase exists in the BCS
through unitary regimes, which is stable only away from the
ground state. 
This is an unusual phase which has both a lower and an upper critical
temperature, as previously found for population imbalanced same mass
Fermi gases \cite{our06}.

\begin{figure} \includegraphics[width=2.8in,clip] {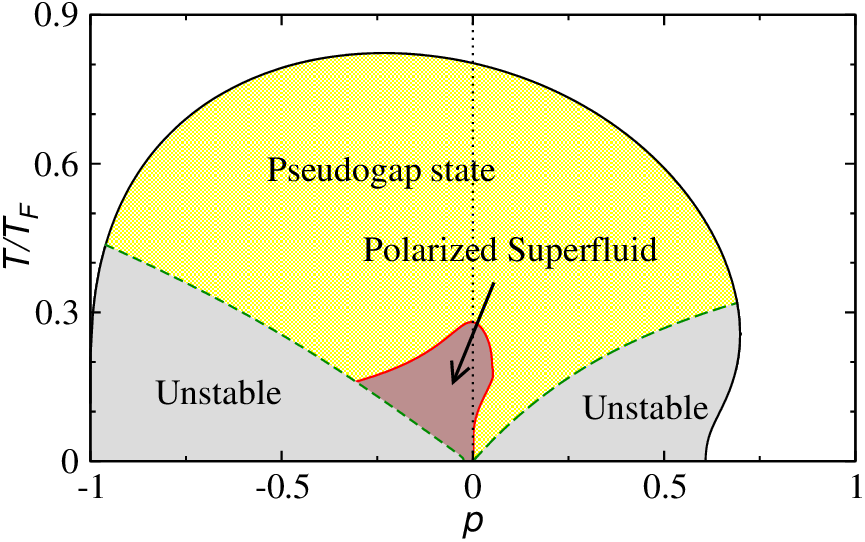} 
\caption{(Color online) Phase diagrams for mixtures of $^{6}$Li and
  $^{40}$K atoms 
  at unitarity. Here $p<0$ when $^{6}$Li is the majority species and
  $p>0$ otherwise. The black solid, red solid, and green dashed lines
  are $T^*$, $T_c$, and the boundary of stable phases,
  respectively. Labeled are polarized superfluid (brown), pseudogap
  state (yellow) and unstable Sarma (grey) phases. The white open space
  is polarized normal Fermi gas. }
\label{fig:unitary} 
\end{figure} 

Since the strongly interacting, unitary regime
% point which corresponds to the onset of an isolated two-body bound
% state
has been the focus of much interest, in Fig.~\ref{fig:unitary} we first
present the phase diagram at unitarity as a function of temperature and
polarization $p$. It shows that uniformly polarized superfluid exists at
low $|p|$ and low but finite $T$ (the dark shaded region). At higher $T$
and higher $|p|$, there is non-superfluid state with a finite excitation
gap (called the pseudogap) before the system becomes essentially
uncorrelated which we refer to as the normal Fermi gas state. However,
at low $T$ and relatively high $|p|$, the Sarma superfluid phase (light
shaded) is found to be unstable. Interestingly, when the heavier species
is the majority (i.e., $p>0$), the stable or unstable paired states
occupy a smaller phase space than the opposite case. 
It should be noted that when phase separation is included, it may cover
the entire unstable region and extend slightly into the pseudogap and
superfluid regions in the phase diagram. In this way the re-entrant
behavior (seen upon a vertical cut a constant low $p$) 
shown in the right hand panel
of Fig.~\ref{fig:unitary} should not survive.

The boundary separating the pseudogap phase and the normal Fermi gas can
be associated with $T^*$, while the line separating the polarized
superfluid and the pseudogap phase corresponds to $T_c$. In a strict
mean field calculation \cite{ParishPRL07} there is no pseudogap so that
the $T_c$ curve falls right on top of the $T^*$ line.
% the pseudogap state would be indistinguishable from the superfluid
% phase.
It is clear, then, that pair fluctuation effects are extremely important
for they greatly reduce the regime where superfluidity
appears. In the equal
mass case, where there is an opportunity to compare with experiments
\cite{MITphase}, a similar reduction in the regime of stable superfluid was
found \cite{RFlong} which was in quite good agreement with the data.
One can see from Fig.~\ref{fig:unitary} that, for both $p<0$ and $p>0$,
while mean-field theory again predicts a large value of the upper
critical polarization (beyond which the superfluid phase cannot exist)
and does not distinguish $T^*$ and $T_c$, 
our theory which includes pairing fluctuations predicts a smaller stable
superfluid regime. Based on evidence in the equal-mass case, our
prediction should be more typical of future experiments.

Another notable feature of the figure is 
that when the heavier species is the majority, a stable homogeneously
polarized superfluid only exists in a very narrow regime of extremely
low polarization. This should serve as an important guide to future
experiments, for it suggests one has to be very careful in order not to miss
%how difficult it will be to observe a
the (homogeneously) polarized superfluid in this case.  It might seem as
though the transition temperatures are much higher in the case of
unequal masses than in the equal mass case. We stress that it is more
meaningful to compare quantities such as $T_c/T^*$ than $T_c/T_F$
because, unlike in the equal mass case, here the energy unit $T_F$ does
not correspond to the Fermi energy of ether species of the atoms.
% is chosen as an interpolation which coincides with that in the
% equal-mass case.
For the equal-mass case $T_c/T^*\approx 0.5$ while $T_c/T^*\approx 0.3$
for a mixture of $^{6}$Li and $^{40}$K atoms at $p=0$.

\begin{figure} \includegraphics[width=2.8in,clip] {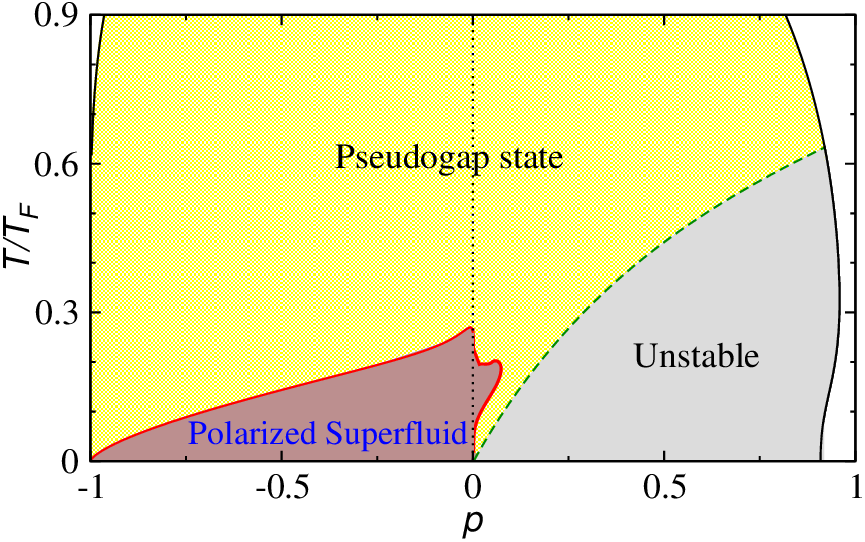} 
\caption{(Color online) Phase diagrams 
for mixtures of $^{6}$Li and $^{40}$K atoms 
at $1/k_{F}a=0.5$. The convention follows that in Fig.~\ref{fig:unitary}}
\label{fig:BEC} 
\end{figure} 

To address how the phase diagram evolves from unitarity to the BEC side
of the Feshbach resonance, 
we present in Fig.~\ref{fig:BEC} the counterpart phase diagrams at
$1/k_F a=0.5$ for the inter-species superfluid. 
It should be clear from the figure that, just as in the previous case,
pair fluctuation effects which allow us (via Eq.~(\ref{eq:1z})) to
distinguish between the gap $\Delta$ and the order parameter
$\Delta_{sc}$, are extremely important as they greatly reduce the 
regime of stable homogeneous superfluidity.

When the heavier species is the majority ($p>0$), there is virtually no
stable polarized superfluid. Again, the re-entrant behavior of the
intermediate temperature superfluid at low $p>0$ in Fig.~\ref{fig:BEC}
should not survive if phase separation is considered.  In contrast, when
the lighter species is the majority ($p<0$), there is no longer an
intermediate temperature superfluid phase. Rather a stable superfluid
can be found for all polarizations and temperatures below the (single)
critical temperature $T_c$.  This behavior can be contrasted with the
equal mass case where, for, e.g., $p = 0.5$ at $T=0$, a stable
superfluid 
cannot be found until deep in the BEC regime, when $1/k_F a>2$.  For a
moderate polarization, say $p=-0.5$, this transition temperature
$T_c\approx 0.18T_F$, which should be accessible in future.

Although we have considered the homogeneous rather than a
trapped case, there are now experimental capabilities for addressing
this phase diagram using tomography \cite{MITtomoimb}. Moreover, we have
previously \cite{RFlong} characterized the changes in the phase diagram
for the equal mass case upon going from the homogeneous to the trapped
situation. We have shown that, while the regimes of stability of
generalized Sarma phases are greatly expanded in a trap, the
characteristic values of the transition temperatures are not
significantly altered.

This work is supported by NSF PHY-0555325 and NSF-MRSEC Grant
0820054. We thank Cheng Chin for helpful conversations.

\vspace*{-1ex}

\bibliographystyle{apsrev} 
%\bibliography{Review2}

\begin{thebibliography}{22}
\bibitem[{exp()}]{expan}
\bibinfo{note}{In contrast to previous work we take higher terms ($a_1
  \Omega^2$) in the T-matrix expansion to avoid numerical complications from
  vanishing $a_0$.}
\end{thebibliography}

\end{document}